\documentclass[final,3p,twocolumn]{elsarticle}

\usepackage{graphicx}

\biboptions{comma,square,numbers,sort&compress}

\journal{Physics Letters B}

\begin{document}


\begin{frontmatter}

\title{Super No-Scale ${\cal F}$-$SU(5)$: A Dynamic Determination of $M_{1/2}$ and $\tan\beta$}

\author[loc_1,loc_2]{Tianjun Li}
\ead{junlt@physics.tamu.edu}

\author[loc_1]{James A. Maxin}
\ead{jmaxin@physics.tamu.edu}

\author[loc_1,loc_3,loc_4]{Dimitri V. Nanopoulos}
\ead{dimitri@physics.tamu.edu}

\author[loc_5]{Joel W. Walker\corref{cor1}}
\ead{jwalker@shsu.edu}
\cortext[cor1]{{\it Corresponding author:} Telephone~+1~(936)~294-4803; Fax~+1~(936)~294-1585}

\address[loc_1]{George P. and Cynthia W. Mitchell Institute for Fundamental Physics,\\
Texas A$\&$M University, College Station, TX 77843, USA }

\address[loc_2]{Key Laboratory of Frontiers in Theoretical Physics, Institute of Theoretical Physics,\\
Chinese Academy of Sciences, Beijing 100190, P. R. China }

\address[loc_3]{Astroparticle Physics Group, Houston Advanced Research Center (HARC),\\
Mitchell Campus, Woodlands, TX 77381, USA}

\address[loc_4]{Academy of Athens, Division of Natural Sciences,\\
28 Panepistimiou Avenue, Athens 10679, Greece }

\address[loc_5]{Department of Physics, Sam Houston State University,
Huntsville, TX 77341, USA }

\begin{abstract}

We study the Higgs potential in No-Scale ${\cal F}$-$SU(5)$, a model built
on the tripodal foundations of the $\cal{F}$-lipped $SU(5)\times U(1)_{\rm X}$
Grand Unified Theory, extra $\cal{F}$-theory derived TeV scale vector-like
particle multiplets, and the high scale boundary conditions
of no-scale supergravity.  $V_{\rm min}$, the minimum of the potential
following radiative electroweak symmetry breaking, is a function
at fixed Z-Boson mass of the universal gaugino boundary mass $M_{1/2}$
and $\tan\beta$, the ratio of Higgs vacuum expectation values.
The so-scale nullification of the bilinear Higgs soft term $B_\mu$
at the boundary reduces $V_{\rm min}(M_{1/2})$ to a one dimensional dependency,
which may be secondarily minimized.  This ``Super No-Scale'' condition
dynamically fixes $\tan\beta$ and $M_{1/2}$ at the local
{\it minimum minimorum} of $V_{\rm min}$.
Fantastically, the walls of this theoretically established
secondary potential coalesce in descent to a striking concurrency with
the previously phenomenologically favored ``Golden Point'' and ``Golden Strip''.

\end{abstract}


\end{frontmatter}


\section{Introduction and Background}

We have recently demonstrated~\cite{Li:2010ws,Li:2010mi} the unique phenomenological
consistency and profound predictive capacity of a model dubbed
No-Scale ${\cal F}$-$SU(5)$, constructed from the merger of the ${\cal F}$-lipped $SU(5)$
Grand Unified Theory (GUT)~\cite{Barr:1981qv,Derendinger:1983aj,Antoniadis:1987dx}, two pairs of
hypothetical TeV scale vector-like supersymmetric multiplets with origins in
${\cal F}$-theory~\cite{Jiang:2006hf,Jiang:2009zza,Jiang:2009za,Li:2010dp,Li:2010rz},
and the dynamically established boundary conditions of no-scale
supergravity~\cite{Cremmer:1983bf,Ellis:1983sf, Ellis:1983ei, Ellis:1984bm, Lahanas:1986uc}.
It appears that the no-scale scenario, particularly vanishing of the Higgs
bilinear soft term $B_\mu$, comes into its own only when applied at an elevated
scale, approaching the Planck mass~\cite{Ellis:2010jb}.  $M_{\cal F}$, the point of the second stage
$SU(5)\times U(1)_{\rm X}$ unification, emerges in turn as a suitable candidate scale
only when substantially decoupled from the primary GUT scale unification
of $SU(3)_C\times SU(2)_L$ via the modification to the renormalization
group equations (RGEs) from the extra ${\cal F}$-theory vector multiplets~\cite{Li:2010ws,Li:2010mi}.

Taking a definition of $M_{\rm V} =$~1~TeV for the new vector-like fields as an elemental 
model feature, we showed~\cite{Li:2010ws} that the viable parameter space consistent
with radiative electroweak symmetry breaking (EWSB), limits on the flavor changing neutral current
$(b \rightarrow s\gamma)$ process and on contributions to the muon anomalous magnetic
moment $(g-2)_\mu$, runs sufficiently perpendicular to both the $B_\mu(M_{\cal F}) = 0$
and centrally observed WMAP 7 cold dark matter (CDM) relic density contours that
the non-trivial mutual intersection is a narrowly confined ``Golden Point'' with a
universal gaugino boundary mass $M_{1/2}$ around 455~GeV, and a ratio
$\tan\beta = 15$ of Higgs vacuum expectation values (vevs).
Insomuch as the collision of top-down model based constraints with bottom-up experimental
data effectively absorbs the final dynamic degree of freedom,
this was labeled a No-Parameter Model.

Advancing from the ``Golden Point'' to the ``Golden Strip''~\cite{Li:2010mi}, we relaxed
the definition of the vector-like mass and studied the impact
of fluctuatating key electroweak reference data ($\alpha_{\rm s}$,$M_{\rm Z}$)
and the top quark mass $m_{\rm t}$ about the error margins.
The most severe variation occurred for $m_{\rm t}$,
via its connection to the large Yukawa coupling essential to radiative EWSB.
We recognized this dependence by effectively treating $m_{\rm t}$ as an additional input,
selecting the appropriate value to restore a vanishing $B_\mu(M_{\cal{F}})$
at each point in the $(M_{1/2}, \tan \beta, M_{\rm V})$ volume.  The $(g-2)_\mu$ and
$(b \rightarrow s\gamma)$ constraints, both at their lower limits, were found to exert
opposing pressures on $M_{1/2}$ due to operation of the former in alignment with,
and the latter in counter-balance of, the Standard Model (SM) leading term. 
Cross cutting by the WMAP CDM measurement completed demarcation of the strip,
running diagonally from about ($M_{1/2},M_{\rm V}$) = (455,1020)~GeV, to (481,691)~GeV,
with $\tan\beta = 15$ independently enforced for the full space.  With
parameterization freedom exhausted, the model was finally required to make a
correlated postdiction for the top quark mass.  The result, $m_{\rm t}=173.0$-$174.4$~GeV,
is in fine accord with the measured value $173.1\pm 1.3$ GeV~\cite{:2009ec}.
The predicted range of $M_{V}$ is testable at the LHC, and the partial lifetime
for proton decay in the leading ${(e|\mu)}^{+} \pi^0 $ channels is
$4.6 \times 10^{34}$ years, testable at the future
Hyper-Kamiokande~\cite{Nakamura:2003hk} and DUSEL~\cite{Raby:2008pd} experiments.

\section{The Super No-Scale Mechanism\label{sct:snsmech}}

In the present work we volunteer a small step backward to emphasize a
giant leap forward.  Having established practical bounds on the vector-like mass, we revert
to a single conceptual universe, ostensibly our own or one of sufficient phenomenological
proximity, with $M_{\rm V} = 1000$~GeV, and $m_{\rm t} = 173.1$~GeV.
Minimization of the Higgs potential with respect to the neutral up-like
and down-like Higgs components $H_u$ and $H_d$ imposes a pair of constraint equations which may be used
to eliminate any two free parameters of the set $M_{1/2}, B_\mu$,
$\tan \beta \equiv  \langle H_u \rangle /  \langle H_d \rangle $, and
the supersymmetry (SUSY) preserving bilinear Higgs mass term $\mu$.
The overall magnitude of the Higgs vev
$v \equiv \sqrt{ \langle H_u \rangle ^2 + \langle H_d \rangle ^2} \simeq 174$~GeV is considered
to be experimentally constrained by measurement of the gauge couplings and Z-Boson mass.
Typically, one will solve for $\mu(M_{\rm Z})$ and $B_\mu(M_{\rm Z})$ in terms of the constrained Higgs vevs
and $\tan \beta$, at fixed $M_{1/2}$.  We consider though that the no-scale boundary condition $B_\mu(M_{\cal{F}}) = 0$
fixes the value of $B_\mu$ at all other scales as well via action of the renormalization group.
Restricting then to just the solution subset for which $B_\mu(M_{\rm Z})$ given by EWSB
stitches cleanly onto that run down under the RGEs from $B_\mu(M_{\cal F})=0$,
$\tan \beta$ (or alternatively $\mu$) becomes an implicit function of the single modulus $M_{1/2}$.
Concretely, we shall consider that the first EWSB constraint absolutely establishes $\mu$, and that the second
gives a line of parameterized solutions for the functional relationship between $M_{1/2}$ and $\tan \beta$.
We therefore distinguish the residual freedom in the dynamic modulus $M_{1/2}$ and parameter $\tan\beta$
by the ability to exert direct influence on the Higgs potential within a single physical parameterization.

The crucial observation is that the minimization of the Higgs potential is therefore at this stage incomplete.
In no-scale supergravity, the specific structure of the K\"ahler potential $K$ leads to a contribution
to the scalar potential which is is zero and flat at tree level, so that the gravitino mass $M_{3/2}$, or by
proportional equivalence $M_{1/2}$, is to be determined dynamically by radiative corrections.
In order to finish specification of the physical vacuum, we must then {\it secondarily}
minimize the Higgs potential with respect to the dependency on $M_{1/2}$, a dependency which is embodied
in the bulk proportionality of the full low energy mass spectrum to this SUSY breaking
parameter~\cite{Ellis:1983sf,Lahanas:1986uc}.  At this locally smallest value of $V_{\rm min}(M_{1/2})$,
which we dub the {\it minimum minimorum}, the dynamic determination $M_{1/2}$ is established.
Moreover, the implicit dependence of the parameter $\tan\beta$ on $M_{1/2}$ means that its value is
also simultaneously provided by the system dynamics.
Henceforth, the imposition of $d V_{\rm min} / d M_{1/2} = 0$ on the Higgs potential will be referred
to as the ``Super No-Scale'' condition.

We emphasize that the justification for this procedure traces back to the fact that the soft SUSY breaking
mass $M_{1/2}$ is related to the F-term of a dynamic modulus.  For example, in the weakly
coupled heterotic $E_8 \times E_8$ string theory, or in M-theory on $S^1 / Z_2$, $M_{1/2}$
is related to the F-term of a K\"ahler modulus $T$.  In string models, there exists
a fundamental question of how any such moduli are to be stabilized.  Thus, the physical
motivation of the Super No-Scale condition is the stabilization of the F-term of the modulus.
Again, for each $M_{1/2}$, we will have an electroweak symmetry breaking vacuum corresponding
to minimization of the scalar Higgs potential.  Among these minima, the {\it minimum minimorum}
is the dynamically preferred locally smallest minimum of the Higgs potential.

We openly recognize that the potential affords an additional dimensionality
along the degree of freedom which has been locked out by
the fixing of $v$, and that minimization with respect to this additional parameter remains a
question of interest.  However, this is a delicate point of ongoing research, and beyond
the scope of the current study.  If one accepts, for the sake of argument,
that the current model fairly represents the physics of our Universe, then current
experimental measurements guarantee that the potential along this direction is indeed bounded,
not running away from the adopted constant value of $v$.  It is therefore only the secondary bounding
along the degree of freedom associated with $M_{1/2}$ which is experimentally unknown to us, and which may be
predicted according to model formulations such as the one here presented.


\section{$\cal{F}$-$SU(5)$ Models}

In the Flipped $SU(5)$ GUTs, 
the gauge group is $SU(5)\times U(1)_{X}$, which embeds in $SO(10)$.
Gauge coupling unification near $10^{16}$~GeV strongly suggests
the existence of a Grand Unified Theory (GUT).
In minimal SUSY $SU(5)$ models there are problems with doublet-triplet
splitting and dimension five proton decay by colored Higgsino exchange~\cite{Antoniadis:1987dx}.
These difficulties are elegantly overcome in Flipped $SU(5)$ GUT
models via the missing partner mechanism~\cite{Antoniadis:1987dx}.
The generator $U(1)_{Y'}$ is defined for fundamental five-plets as
$-1/3$ for the triplet members, and $+1/2$ for the doublet.
The hypercharge is given by $Q_{Y}=( Q_{X}-Q_{Y'})/5$.
There are three families of Standard Model (SM) fermions, a pair of ten-plet
Higgs for breaking the GUT symmetry, and a pair
of five-plet Higgs for EWSB. 

Historically, the first flipped F-theory $SU(5)$ GUT was constructed in
Ref.~\cite{Beasley:2008kw}, and further aspects of flipped $SU(5)$ F-theory GUTs have been
considered in \cite{Chen:2010tp, Chung:2010bn, Kuflik:2010dg}.
We introduce in addition, vector-like particle multiplets, derived likewise in the context of
F-theory model building~\cite{Jiang:2006hf}, to address the ``little hierarchy''
problem, altering the beta coefficients of the renormalization group to dynamically elevate the secondary 
$SU(5)\times U(1)_{\rm X}$ unification at $M_{\cal F}$ to near the Planck
scale, while leaving the $SU(3)_C\times SU(2)_L$ unification at $M_{32}$
close to the traditional GUT scale.  In other words,
one obtains true string-scale gauge coupling unification in 
free fermionic string models~\cite{Jiang:2006hf,Lopez:1992kg} or
the decoupling scenario in F-theory models~\cite{Jiang:2009zza,Jiang:2009za}.
To avoid a Landau pole for the strong
coupling constant, we are restricted around the TeV scale
to one of the following two multiplet sets~\cite{Jiang:2006hf}.
\begin{eqnarray}
\hspace{-.3in}
& \left( {XF}_{\mathbf{(10,1)}} \equiv (XQ,XD^c,XN^c),~{\overline{XF}}_{\mathbf{({\overline{10}},-1)}} \right)& \nonumber \\
\hspace{-.3in}
& \left( {Xl}_{\mathbf{(1, -5)}},~{\overline{Xl}}_{\mathbf{(1, 5)}}\equiv XE^c \right)& \label{z1z2}
\end{eqnarray}
Prior, $XQ$, $XD^c$, $XE^c$, $XN^c$ have the same quantum numbers as the
quark doublet, right-handed down-type quark, charged lepton, and
neutrino, respectively.  We have argued~\cite{Li:2010mi} that the eminently
feasible near-term detectability of these hypothetical fields in collider experiments,
coupled with the distinctive flipped charge assignments of the multiplet structure,
represents a smoking gun signature for Flipped $SU(5)$, and have thus coined the term
{\it flippons} to collectively describe them.
Immediately, our curiosity is piqued by the announcement~\cite{Abazov:2010ku}
of the D\O~collaboration that vector-like quarks have been excluded up to
a bound of 693~GeV, corresponding to the lower edge of our golden strip.
We here consider only the $Z2$ set, although discussion for the $Z1$ set,
if supplemented by heavy threshold corrections, will be similar.


\section{No-Scale Supergravity}

The Higgs boson, being a lorentz scalar,
is not stable in the SM against quadratic quantum mass corrections
which drive it toward the dominant Planck scale, some
seventeen orders of magnitude above the value required for consistent 
EWSB.  Supersymmetry naturally solves this fine tuning problem
by pairing the Higgs with a chiral spin-$1/2$ ``Higgsino'' partner field, and
following suit with a corresponding bosonic (fermionic) superpartner for all
fermionic (bosonic) SM fields, introducing the full set of quantum counter terms.
Localizing the supersymmetry algebra, which includes the generator of
spacetime translations (the momentum operator),
induces general coordinate invariance, producing the
supergravity (SUGRA) theories.

Since we do not observe mass degenerate superpartners for the known SM fields,
SUSY must itself be broken around the TeV scale.
In the traditional framework, supersymmetry is broken in 
the hidden sector, and the effect is 
mediated to the observable sector via gravity or gauge interactions.
In GUTs with minimal gravity mediated supersymmetry breaking, called mSUGRA,
one can fully characterize the supersymmetry breaking
soft terms by four universal parameters
(gaugino mass $M_{1/2}$, scalar mass $M_0$, trilinear coupling $A$, and
the low energy ratio $\tan\beta$),
plus the sign of the Higgs bilinear mass term $\mu$.

No-Scale Supergravity was proposed~\cite{Cremmer:1983bf,Ellis:1983sf, Ellis:1983ei, Ellis:1984bm, Lahanas:1986uc}
to address the cosmological flatness problem.
It may be verified for the simple example K\"ahler potential
\begin{eqnarray}
K &=& -3\,\ln\,( T+\overline{T}-\sum_i \overline{\Phi}_i \Phi_i)~,~
\label{NS-Kahler}
\end{eqnarray}
where $T$ is a modulus field and $\Phi_i$ are matter fields,
that the no-scale boundary conditions $M_0=A=B_{\mu}=0$ are
enforced automatically, while $M_{1/2}>0$ is allowed,
as is indeed required for SUSY breaking.
All low energy scales are dynamically generated by quantum corrections,
{\it i.e.}~running under the RGEs, to the classically flat potential.
Additionally, the tree level vacuum energy vanishes automatically.
The fiercely reductionist no-scale picture moreover inherits
an associative weight of motivation from its robustly generic
and natural appearance in string based constructions.

The simple from of Eq.~(\ref{NS-Kahler}) has been independently derived in both
weakly coupled heterotic $E_8 \times E_8$ string theory~\cite{Witten:1985xb} and
for strong coupling, in the leading order compactification of M-theory on
$S^1/Z_2$~\cite{Li:1997sk}.  In both cases, the Yang-Mills fields span a ten
dimensional space-time.  However, this potential is not obtained directly out of
F-theory, as represented for example by the strong coupling lift from Type
IIB intersecting D-brane model building with D7- and
D3-branes~\cite{Beasley:2008dc,Beasley:2008kw, Donagi:2008ca, Donagi:2008kj},
where the Yang-Mills fields on the D7-branes occupy an eight dimensional space-time.
Nevertheless, it is certainly possible in principle to
calculate a gauge kinetic function, Kahler potential and superpotential
in the context of Type IIB interecting D-brane model building, and
the F-theory could thus admit a more general definition
of no-scale supergravity, as realized by a K\"ahler potential like
\begin{eqnarray}
K &=& -\ln(S + \overline{S}) -\ln(T_1 + \overline{T}_1)
\nonumber \\
&& -\ln(T_2 + \overline{T}_2) -\ln(T_3 + \overline{T}_3) \, ,
\label{Kahler2}
\end{eqnarray}
where only three of the moduli fields $S$ and $T_i$ may yield non-zero F-terms.

However, the ${\cal F}$-$SU(5)$ type models under discussion have been constructed locally
in F-theory~\cite{Jiang:2009zza, Jiang:2009za}, and without a corresponding consistent global
construction, we do not know the concrete K\"ahler potential of the SM fermions and Higgs fields,
and cannot by this means explicitly calculate the supersymmetry breaking scalar masses and
trilinear soft terms.  Essentially then, we aim to study an F-theory
{\it inspired} variety of low energy SUSY phenomenology, remaining agnostic
as to the details of the K\"ahler structure.  By studying the simplest no-scale
supergravity, we may still however expect to encapsulate the correct leading order behavior.
Should the favorable qualitative phenomenology of this lowest order analysis prove persistent,
our future attention will be directed toward quantitatively specific no-scale supergravity generalizations.


\begin{figure*}[htf]
        \centering
        \includegraphics[width=0.55\textwidth]{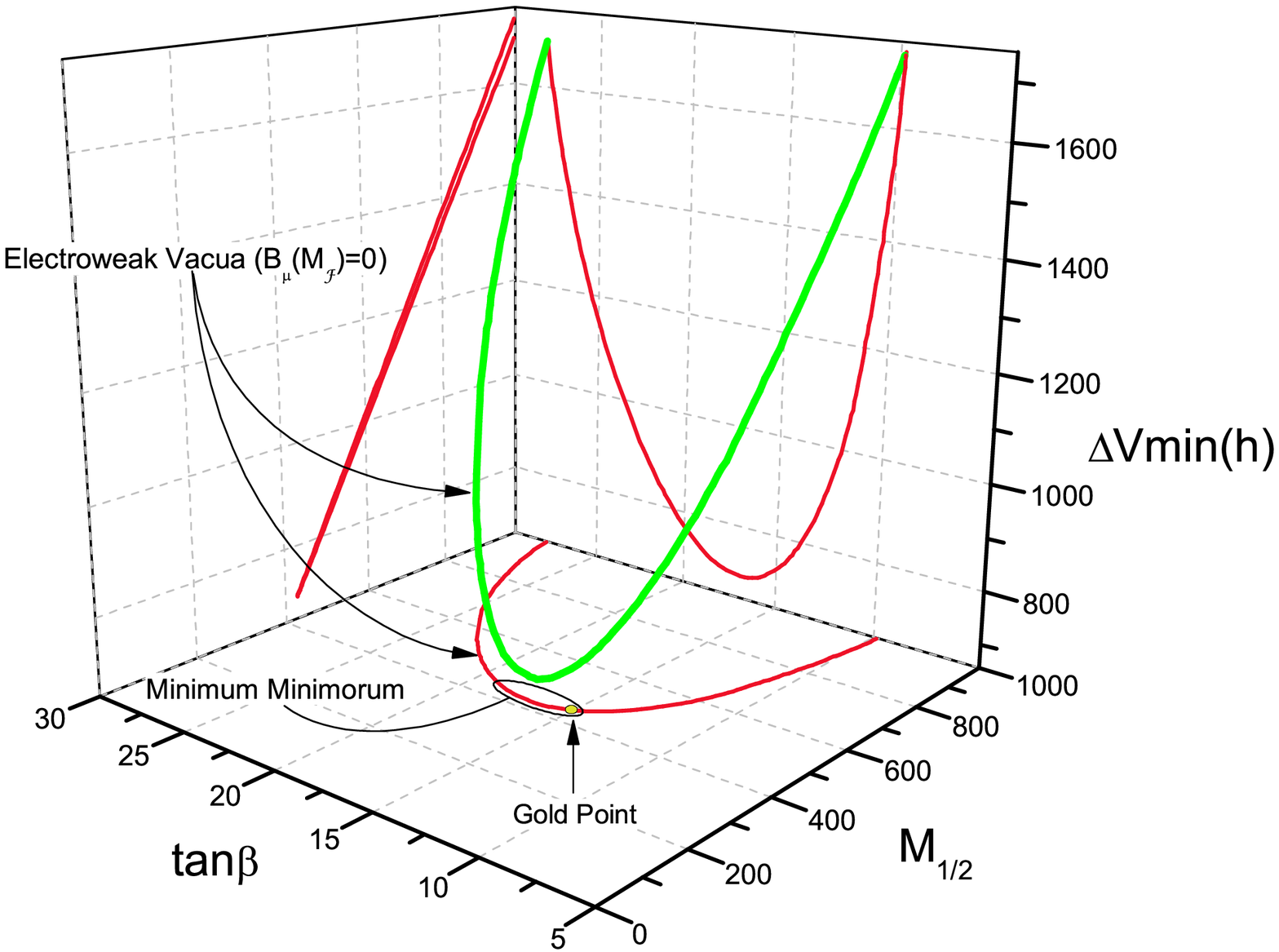}
        \hspace{0.00\textwidth}
        \includegraphics[width=0.40\textwidth]{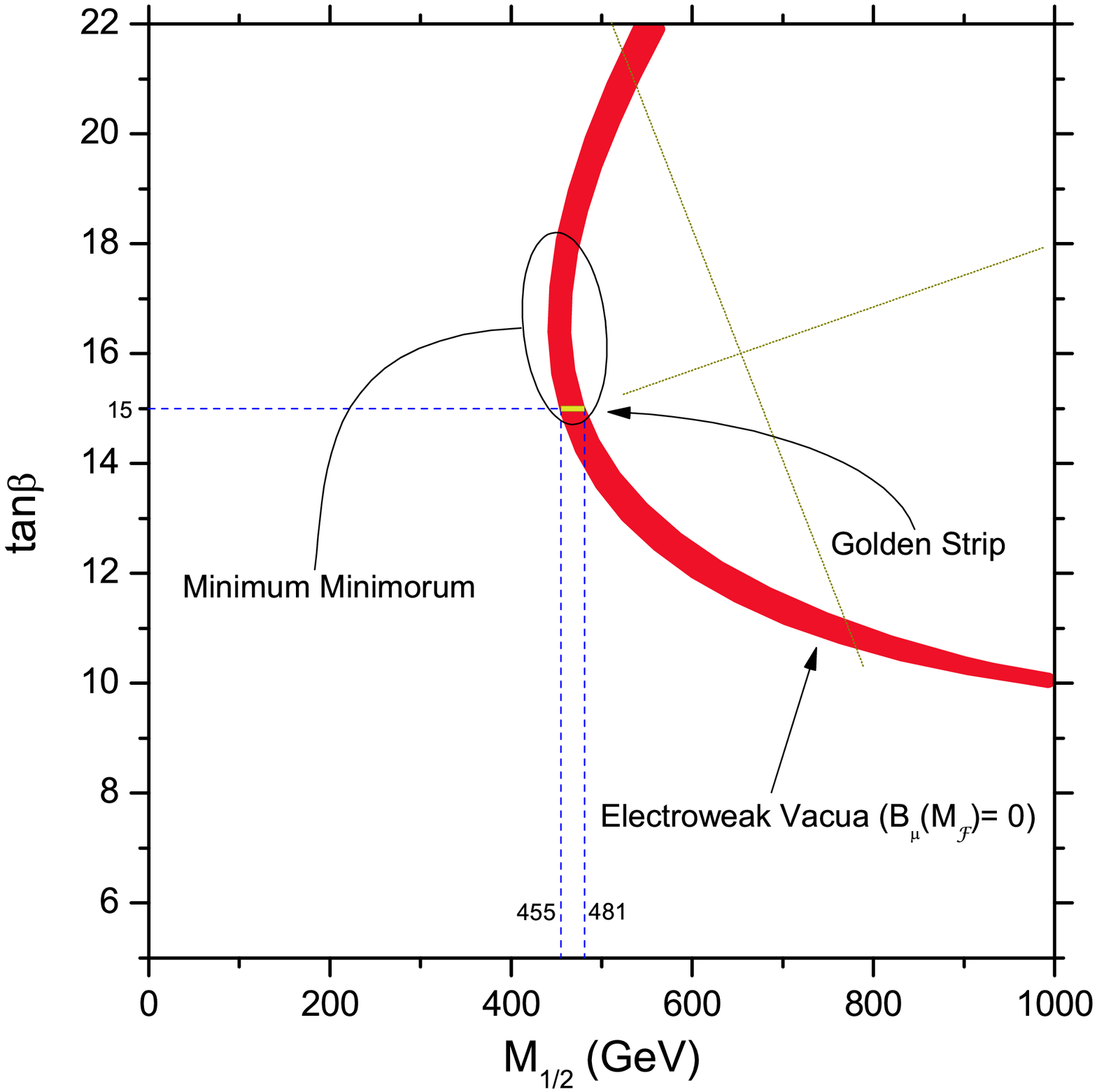}
        \caption{
$A)$ The {\it minimum} $V_{\rm min}$ of the Higgs effective potential
or more precisely, the signed fourth root of the energy density
${\rm Sign}(V_{\rm min}) \times {\vert V_{\rm min} \vert }^{1/4}$,
is plotted (green curve, GeV) as a function of $M_{1/2}$ (GeV) and $\tan \beta$, 
emphasizing proximity of the ``golden point'' of Ref.~\cite{Li:2010ws}
to the dynamic region of the $V_{\rm min}$ {\it minimorum}.
$B)$ The projection onto the ($M_{1/2}$,tan$\beta$) plane is further
detailed in the second figure, expanding to span the boundary cases of
the Ref.~\cite{Li:2010mi} ``golden strip''.  The symmetry axis of the $B_{\mu} = 0$
parabola is rotated slightly above the $M_{1/2}$ axis.
        }
        \label{fig:vmin}
\end{figure*}


\section{The Higgs Minimum Minimorum}

We now proceed to specifically implement, within the context of the $\cal{F}$-$SU(5)$
construction, the Super No-Scale mechanism described in Section~(\ref{sct:snsmech}).
Again, for a given Higgs vev, {\it i.e.}~for a fixed Z-Boson mass, we establish $\tan\beta$,
by application of the two EWSB consistency conditions, to be an implicit function of the
universal gaugino boundary mass $M_{1/2}$, along a continuous string of minima of the
broken Higgs potential $V_{\rm min}$, which are likewise labeled by their value of $M_{1/2}$.
It is with respect to this line of solutions that we seek to establish a local secondary
{\it minimum minimorum} of the Higgs potential $V_{\rm min} (M_{1/2})$.

We employ an effective Higgs potential in the 't Hooft-Landau gauge and the $\overline{\rm DR}$
scheme, given summing the following neutral tree $(V_0)$ and one loop $(V_1)$ terms.
\begin{eqnarray}
\hspace{-.3in}
V_0 &=& (\mu^2 + m_{H_u}^2) (H_u^0)^2 + (\mu^2 + m_{H_d}^2) (H_d^0)^2 \nonumber \\
\hspace{-.3in}
&& -2 \mu B_{\mu} H_u^0 H_d^0 + {\frac{g_2^2 + g_Y^2}{8}} \left[(H_u^0)^2-(H_d^0)^2\right]^2 \nonumber \\
\hspace{-.3in}
V_1 &=&  \sum_i {\frac{n_i}{64\pi^2}} m_i^4(\phi) \left( {\rm ln}{\frac{m_i^2(\phi)}{Q^2}} -{\frac{3}{2}} \right) \label{higgs_v1}
\end{eqnarray}
Prior, $m_{H_u}^2$ and $m_{H_d}^2$ are the soft SUSY
breaking masses of the Higgs fields $H_u$ and $H_d$,
$g_2$ and $g_Y$ are the gauge couplings of $SU(2)_L$ and
$U(1)_Y$, $n_i$ and $m_i^2(\phi)$ are the degree of freedom and mass
for $\phi_i$, and $Q$ is the renormalization scale. 
In particular, the soft breaking parameters $m^2_{H_u}$ and $m^2_{H_d}$ are not free parameters,
but rather functions of the universal gaugino boundary mass $M_{1/2}$, run down to the point of electroweak
symmetry breaking under the renormalization group.
We include the complete Minimal Supersymmetric Standard Model
 (MSSM) contributions to one loop,
following Ref.~\cite{Martin:2002iu}, although the result is phenomenologically
identical accounting only the leading top and partner stop terms.
Since the minimum of the electroweak (EW) Higgs potential
$V_{\rm min}$ depends implicitly on $M_{3/2}$,  the gravitino mass is
determined by the Super No-Scale condition $dV_{\rm min}/dM_{3/2}=0$.
Being, however, that $M_{1/2}$ is proportional to $M_{3/2}$,
it is equivalent to employ $M_{1/2}$ directly as our modulus parameter,
as previously described.  All other SUSY breaking soft terms will
subsequently be derived from this single dynamically determined value.

Factors explicit within the potential are obtained from 
our customized extension of the {\tt SuSpect 2.34}~\cite{Djouadi:2002ze}
codebase, including a self-consistency
assessment~\cite{Li:2010ws} on $B_\mu = 0$.
We apply two-loop RGE running for the SM gauge couplings,
and one-loop running for the SM fermion Yukawa couplings, 
$\mu$ term and soft terms.

Studying $V_{\rm min}$ generically in the $(M_{1/2}, \tan \beta)$ plane, no point of secondary
minimization is readily apparent in the strong, roughly linear, downward trend with
respect to $M_{1/2}$ over the region of interest.  However, the majority of the plane
is not in physical communication with our model, disrespecting the fundamental
$B_\mu = 0$ condition.  Isolating only the compliant $B_\mu = 0$ contour within this surface,
a parabola is traced, the nadir of which is in excellent agreement with 
our original golden point, as shown in Fig.~(\ref{fig:vmin}A).
Restoring parameterization freedom to $(M_{\rm V},m_{\rm t})$, we may scan across the corresponding
golden point of each nearby universe variant, reconstructing in their union the previously
advertised golden strip, as in Fig.~(\ref{fig:vmin}B).  Notably, the theoretical
restriction on $\tan\beta$ remains stable against variation in these parameters,
exactly as its experimental counterpart. 

We find it quite extraordinary that the phenomenologically preferred region
rests precisely at the curve's locus of symmetric inflection.
Note in particular that it is the selection of the parabolic $B_\mu=0$ contour
out of the otherwise uninteresting $V_{\rm min}(M_{1/2},\tan \beta)$ inclined surface
which allows a clear {\it minimum minimorum} to be established.
We reiterate that consistency of the dynamically positioned
$M_{1/2}$ and $\tan\beta$ with the golden strip
implies broad consistency with all current experimental data,
within the resolution of the methodology and numerical tools employed.

A strongly linear relationship is observed between the SUSY
and EWSB scales with $M_{\rm EWSB} \simeq 1.44~M_{1/2}$, such that a corresponding 
parabolic curve may be visualized.
There is a charged stau LSP for $\tan\beta$ from 16 to 22,
and we connect points with correct EWSB smoothly on the plot in this region.
If $\tan\beta$ is larger than 22, the stau is moreover
tachyonic, so properly we must restrict all analysis to $\tan\beta \leq 22$.

\section{Additional Phenomenology}

\begin{table*}[ht]
  \small
        \centering
        \caption{Sparticle and Higgs spectrum (in GeV) for the $M_{1/2} = 455$~GeV
and $\tan \beta = 15$ ``Golden Point'' of Ref.~(\cite{Li:2010ws}).
Here, $\Omega_{\chi} = 0.114$, $\sigma_{SI} = 1.9 \times 10^{-10}$~pb, and
$\left\langle \sigma v \right\rangle_{\gamma\gamma} = 1.7 \times 10^{-28}~{\rm cm}^{3}{\rm /s}$.
The central prediction for the $p \!\rightarrow\! {(e\vert\mu)}^{\!+}\! \pi^0$
proton lifetime is $4.6 \times 10^{34}$~years.}
                \begin{tabular}{|c|c||c|c||c|c||c|c||c|c||c|c|} \hline
    $\widetilde{\chi}_{1}^{0}$&$95$&$\widetilde{\chi}_{1}^{\pm}$&$185$&$\widetilde{e}_{R}$&$150$&$\widetilde{t}_{1}$&$489$&$\widetilde{u}_{R}$&$951$&$m_{h}$&$120.1$\\ \hline
    $\widetilde{\chi}_{2}^{0}$&$185$&$\widetilde{\chi}_{2}^{\pm}$&$825$&$\widetilde{e}_{L}$&$507$&$\widetilde{t}_{2}$&$909$&$\widetilde{u}_{L}$&$1036$&$m_{A,H}$&$920$\\ \hline

    $\widetilde{\chi}_{3}^{0}$&$820$&$\widetilde{\nu}_{e/\mu}$&$500$&$\widetilde{\tau}_{1}$&$104$&$\widetilde{b}_{1}$&$859$&$\widetilde{d}_{R}$&$992$&$m_{H^{\pm}}$&$924$\\ \hline
    $\widetilde{\chi}_{4}^{0}$&$824$&$\widetilde{\nu}_{\tau}$&$493$&$\widetilde{\tau}_{2}$&$501$&$\widetilde{b}_{2}$&$967$&$\widetilde{d}_{L}$&$1039$&$\widetilde{g}$&$620$\\ \hline
                \end{tabular}
                \label{tab:masses}
\end{table*}

No-Scale $\cal{F}$-$SU(5)$ features, quite stably, the distinctive mass hierarchy
$m_{\tilde{t}} < m_{\tilde{g}} < m_{\tilde{q}}$
of a light stop and gluino, both comfortably lighter than all other
squarks.  Typical ballpark mass values consistent with the dynamic
determination of $M_{1/2} \sim 450$~GeV
are $m_{\tilde{t}} \sim 500$~GeV, $m_{\tilde{g}} \sim 625$~GeV, and
$m_{\tilde{q}} \sim 1000$~GeV. The lightest neutralino, which is $\sim 99.8\%$ Bino,
may feature a mass somewhat less than $100$~GeV.
For direct comparison, we reprint the detailed spectrum of the original
``Golden Point'' of Ref.~\cite{Li:2010ws} in Table~(\ref{tab:masses}).

We suggest that the spectrum so described thus far survives the advancing detection limits
being posted by early LHC results, which we further point out are typically tuned to the
mSUGRA/CMSSM context, often also with particular assumptions applied to $\tan \beta$.
However, the margin of escape may be narrow, even for the meager $35~{\rm pb}^{-1}$
of integrated luminosity heretofore described.  Specifically, Figure~(2) of Ref.~\cite{daCosta:2011qk}
seems to imply a $95 \%$ lower exclusion boundary of slightly more than $1$~TeV for
gluino masses in our favored range.  The approach taken by this example analysis does
at least take a step toward probing No-Scale $\cal{F}$-$SU(5)$ by the claim of model
independence from the mSUGRA/CMSSM orthodoxy, but there are several peculiar assumptions made
with regards to the spectrum that suggest the prudence of a certain circumspection in interpretation of 
any quoted bottom line results.  In particular, the lightest neutralino is made massless,
and all SUSY fields besides the gluino and the first two squark generations, {\it i.e.}~all sleptons,
all Higgs, all other neutralino components and the third generation of squarks, are decoupled by assignment
of an ultra-heavy 5~TeV mass.  A closer inspection of the data files published by the ATLAS
collaboration along with the cited report confirms that all decay modes are eliminated besides 
those to the massless neutralino plus hadronic jets or leptons.

We maintain some ever present anticipation that the discovery of supersymmetry at the LHC could be imminent,
a sentiment which an optimistic reading of the early reports from ATLAS and CDF might be taken to reinforce.
We should remark, however, that relaxation of the fixed $M_{\rm V}$ and $m_{\rm t}$ mass
values adopted here for simplicity and concreteness will allow the migration, if necessary, to a somewhat
heavier spectrum.  This may be accomplished without wholesale rejection of the underlying model
(No-Scale $\cal{F}$-$SU(5)$) or method of analysis (the Super No-Scale mechanism) which
have been our focus in the present work.  We defer for future work a comprehensive mapping
of such alternative configurations, which in their union compose the complete viable model space.


\section{The Gauge Hierarchy Problem}

The ``gauge hierarchy problem'' represents, in actuality, the clustering of multiple
related difficulties into a single amalgamation, rather than a single isolated problem
with a correspondingly isolated solution. 
Not only must we explain stabilization of the EW scale against quantum corrections,
but we must also explain why this scale and TeV-sized SUSY breaking
soft-terms are ``initially'' positioned so far below the Planck mass.
These latter components of the gauge hierarchy problem are the more subtle.
In their theoretical pursuit, we do not though feign ignorance of
established experimental boundaries, taking the phenomenologist's
perspective that pieces fit already to the puzzle stipulate a partial contour
of those yet to be placed.  Indeed, careful knowledge of
precision EW scale physics, including the strong and
electromagnetic couplings, the Weinberg angle and $M_{\rm Z}$ 
are required even to run the one loop RGEs.  In the second
loop, one requires also {\it minimally} the leading top quark Yukawa coupling, as
deduced from $m_{\rm t}$, and the overall magnitude of the Higgs vev $v$,
established in turn from measurement of the effective Fermi coupling,
or from $M_{\rm Z}$ and the electroweak couplings.

Reading the RGEs up from $M_{\rm Z}$,
we take unification of the gauge couplings as evidence of a GUT.
Reading them in reverse from a point of high energy unification, we take the
heaviness of the top quark, via its large Yukawa coupling, to dynamically
drive the term $M_{\rm H_u}^2 +\mu^2$ negative, triggering spontaneous
collapse of the tachyonic vacuum, {\it i.e.}~radiative electroweak symmetry breaking.
As we have elaborated in Section(\ref{sct:snsmech}), the minimization of this potential with
respect to the neutral components of $H_u$ and $H_d$ at fixed Z-Boson mass allows one
to absolutely establish a numerical value for $\mu$, in addition to a line of 
of continuously parameterized solutions for the functional relationship between $M_{1/2}$ and $\tan \beta$.

Strictly speaking though, we must recognize that having effectively exchanged input of the Z-mass
for a constraint on $\mu (M_{\cal F})$, we dynamically establish the SUSY breaking soft term
$M_{1/2}$ and $\tan \beta$ {\it within} the electroweak symmetry breaking
vacua, {\it i.e.}~with fixed $v\simeq 174$~GeV.  By employing only values of $\mu$ consistent
with the physically constrained Higgs vev, the current construction does not then intrinsically
address the $\mu$ problem, {\it i.e.}~the reason for the proximity of the SUSY preserving Higgs mass
parameter $\mu$ to the electroweak scale and the soft SUSY breaking mass term $M_{1/2}$.
This problem is however ubiquitous to all supersymmetric model constructions, and there is no reason to prevent a
parallel embedding of the usual proposals for addressing the $\mu$ problem alongside the Super No-Scale mechanism.
Likely candidates for the required suppression relative to the Planck scale would include the invocation 
of powers of F-term vevs $\langle F \rangle/M_{\rm Pl}$ via the Giudice-Masiero mechanism~\cite{Giudice:1988yz},
or the introduction of a SM singlet Higgs field as in the next-to-minimal supersymmetric standard
model (NMSSM), or as a final example, the consideration of an anomalous $U(1)_{\rm A}$ gauge symmetry
to be realized out of a string theoretic model building approach~\cite{Babu:2002ic}.

Acknowledging that we have not here fundamentally explained the TeV-scale correlation of $\mu$ and $M_{\rm V}$
to the modulus $M_{1/2}$, we are nevertheless content to justify the values employed by the success of the
globally consistent picture which they facilitate.  In any event, a clear conceptual distinction should be maintained
between the simple parameters $\mu$ and $M_{\rm V}$ and the string theoretic modulus $M_{1/2}$, the latter being 
uniquely eligible for dynamic stabilization under application of the Super No-Scale mechanism.
The current proposal may reach somewhat farther though, than even it first appears.
Having predicted $M_{\cal F}$ as an output scale near
the reduced Planck mass, we are licensed to invert the solution,
taking $M_{\cal F}$ as a high scale {\it input} and dynamically
address the gauge hierarchy through the standard story of radiative electroweak
symmetry breaking.  This proximity to the elemental high scale of
(consistently decoupled) gravitational physics, arises because of the
dual flipped unification and the perturbing effect of the TeV multiplets, 
and is not motivated in standard GUTs.

Operating the machinery of the RGEs in reverse,
we may transmute the low scale $M_{\rm Z}$ for the high scale $M_{\cal F}$,
emphasizing that the fundamental dynamic {\it correlation} is that of 
the {\it ratio} $M_{\rm Z}/M_{\cal F}$, taking either as our input yardstick according to taste.
For fixed $M_{\cal F} \simeq 7\times10^{17}$~GeV, in a single breath we receive
the order of the electroweak scale, the Z-mass, the Higgs bilinear coupling $\mu$ and the
Higgs vevs, all while dynamically tethering this derived scale to the soft SUSY breaking parameter $M_{1/2}$ via
the action of the secondary minimization $d V_{\rm min} / d M_{1/2} = 0$.  All other dependent dimensional
quantities, including the full superparticle mass spectrum likewise then fall into line.
It is in this sense that the Super No-Scale mechanism, as applied to the
present No-Scale $\cal{F}$-$SU(5)$ construction, may contribute to an understanding of
the issues composing the gauge hierarchy problem.


\section{Conclusion}

In this Letter, we have explored the
Super No-Scale condition, that being the dynamic
localization of the {\it minimum minimorum} of the Higgs potential,
{\it i.e.}~a locally smallest value of $V_{\rm min}(M_{1/2})$,
such that both $\tan\beta$ and $M_{1/2}$ are determined.
The stabilized supersymmetry breaking and electroweak scales may both
be considered as dependent output of this construction, thus
substantively addressing the gauge hierarchy problem in the
No-Scale ${\cal F}$-$SU(5)$ context.
We have furthermore demonstrated the striking concurrence of this
theoretical result with the previously phenomenologically favored
``golden point'' and ``golden strip''.

By comparison, the standard MSSM construction seems a hoax, requiring horrendous fine
tuning to stabilize if viewed as a low energy supergravity limit, and moreover
achieving TeV scale EW and SUSY physics as a simple shell game by manual 
selection of TeV scale boundaries for the soft terms $M_{1/2}$, $M_0$, and $A$.
It is remarkable that despite featuring more freely tunable parameters, these constructions are
finding it increasingly difficult to reconcile their phenomenology with early LHC data.


\section{Acknowledgments}

This research was supported in part 
by  the DOE grant DE-FG03-95-Er-40917 (TL and DVN),
by the Natural Science Foundation of China 
under grant No. 10821504 (TL),
and by the Mitchell-Heep Chair in High Energy Physics (TL).

\bibliographystyle{model1-num-names}

\bibliography{bibliography.bib}

\end{document}